\documentclass[12pt]{article}

\newcommand{\half}{ {\textstyle\frac{1}{2}} }

\newcommand{\re}{\mbox{Re}\,}
\newcommand{\im}{\mbox{Im}\,}

\newcommand{\ms}{\mskip 1.5mu}

\begin{document}

\begin{center}
{\bfseries DISPERSION REPRESENTATIONS FOR \\ HARD EXCLUSIVE REACTIONS}%
\,\footnote{Talk presented by D.I.\ at the 12th Workshop on High
  Energy Spin Physics (DSPIN-07), Dubna, Russia, September 3--7, 2007}

\vskip 5mm M.~Diehl$^{1}$ and \underline{D.Yu.~Ivanov}$^{2\dag}$

\vskip 5mm {\small (1) {\it Theory Group, Deutsches
Elektronen-Synchroton DESY, 22603 Hamburg, Germany
}\\
(2) {\it Sobolev Institute of Mathematics, 630090 Novosibirsk,
Russia
}\\
$\dag$ {\it E-mail: d-ivanov@math.nsc.ru }}
\end{center}

\vskip 5mm
\begin{abstract}
  A number of hard exclusive scattering processes can be described in
  terms of generalized parton distributions (GPDs) and perturbative
  hard-scattering kernels.  Both the physical amplitude and the
  hard-scattering kernels fulfill dispersion relations.  We show that
  their consistency at all orders in perturbation theory is guaranteed
  if the GPDs satisfy certain integral relations.  These relations are
  fulfilled thanks to Lorentz invariance.
\end{abstract}

\vskip 8mm


\section{Introduction}

For hard exclusive processes that can be calculated using collinear
factorization, one may write down dispersion relations both for the
physical process and for the parton-level subprocess.  The question of
consistency between both representations turns out to be nontrivial.
Important progress has recently been reported in
\cite{Teryaev:2005uj}, where it was shown that this consistency is
ensured by Lorentz invariance in the form of the polynomiality
property for generalized parton distributions (GPDs).  The studies in
\cite{Teryaev:2005uj} were carried out using the Born-level
approximation of the hard-scattering subprocess.  In particular, they
showed that to this accuracy not only the imaginary but also the real
part of the process amplitude can be represented in terms of GPDs
$F(x,\xi,t)$ along the line $x=\xi$ in the $x$--$\xi$ plane. It is
natural to ask how the situation changes when including radiative
corrections to the hard-scattering kernel.  Here we discuss dispersion
representations for hard exclusive processes to all orders in
perturbation theory, generalizing the leading-order results derived
for the unpolarized quark GPDs in \cite{Teryaev:2005uj}. More details,
as well as results for polarized quarks and for gluons, where special
issues arise, can be found in our journal publication
\cite{Diehl:2007jb}.


\section{Dispersion relations}

The exclusive processes we discuss here are deeply virtual Compton
scattering (DVCS) and light meson production,
\begin{equation}
  \label{procs}
\gamma^*(q) + p(p) \to \gamma(q') + p(p') \,,\qquad
\gamma^*(q) + p(p) \to M(q') + p(p') \,,
\end{equation}
where four-momenta are indicated in parentheses.  Since the processes
in (\ref{procs}) involve particles with nonzero spin, the appropriate
quantities for discussing dispersion relations are invariant
amplitudes, which have simple analyticity and crossing properties. An
explicit decomposition for Compton scattering can be found in
\cite{Belitsky:2001ns}.

We use the Mandelstam variables $s=(p+q)^2$, $t=(p-p')^2$,
$u=(p-q')^2$, and consider an invariant amplitude ${\cal
  F}^{[\sigma]}(\nu,t)$ with definite signature $\sigma$ under $s
\leftrightarrow u$ crossing,
\begin{equation}
  \label{crossing}
{\cal F}^{[\sigma]}(-\nu,t)
  = \sigma {\cal F}^{[\sigma]}(\nu,t) \,,
\end{equation}
where $2\nu = s-u$.  At $t \le 0$ the imaginary part of the amplitude
is due to the $s$-channel discontinuity for $\nu>0$ and to the
$u$-channel discontinuity for $\nu<0$. The fixed-$t$ dispersion
relation with one subtraction reads
\begin{eqnarray}
  \label{disp-sub}
\lefteqn{
\re {\cal F}^{[\sigma]}(\nu,t) - \re {\cal F}^{[\sigma]}(\nu_0,t)
}
\nonumber \\
 &=&
\frac{1}{\pi} \int_{\nu_{th}}^{\infty} d\nu'\,
  \im{\cal F}^{[\sigma]}(\nu',t)
  \left[ \frac{1}{\nu'-\nu} + \sigma \frac{1}{\nu'+\nu}
       - \frac{1}{\nu'-\nu_0} - \sigma \frac{1}{\nu'+\nu_0} \right] ,
\end{eqnarray}
where $\nu_{th}$ is the value of $\nu$ at threshold. Its validity
requires
\begin{equation}
  \label{no-sub}
\nu^{-2} {\cal F}^{[+]}(\nu,t) \to 0 \,, \qquad\qquad
\nu^{-1} {\cal F}^{[-]}(\nu,t) \to 0 \,.
\end{equation}
for $|\nu| \to\infty$.  We consider dispersion relations for the
processes (\ref{procs}) in the Bjorken limit of large $-q^2$ at fixed
$q^2/\nu$ and $t$.  It is useful to trade $\nu$ for the scaling
variable
\begin{equation}
  \label{xi-def}
\xi = -\frac{(q+q')^2}{2\ms (p+p')\cdot (q+q')}
    = -\frac{q^2}{s-u} = -\frac{q^2}{2\nu} \,,
\end{equation}
where we have neglected $q'^2$ and $t$ compared with $q^2$ in the
numerator.  The factorization theorems state that in the Bjorken limit
certain invariant amplitudes become dominant and can be written as
convolutions of partonic hard-scattering kernels with quark or gluon
GPDs. We discuss the contribution of unpolarized quark distributions
$F^q = \{ H^q, E^q \}$ to the leading invariant amplitudes for DVCS or
meson production,
\begin{equation}
  \label{fact-form}
{\cal F}^{q [\sigma]}(\xi,t) = \int_{-1}^1 dx\, \frac{1}{\xi}\,
  C^{q [\sigma]}\Bigl( \frac{x}{\xi} \Bigr)\,
  F^{q}(x,\xi,t)
\end{equation}
where for brevity we do not display the dependence of ${\cal F}^{q
  [\sigma]}$ and $C^{q [\sigma]}$ on $q^2$.  The hard-scattering
kernel satisfies the symmetry relation
\begin{equation}
  \label{kernel-symm}
  C^{q [\sigma]}(-x/\xi)
= -\sigma\ms C^{q [\sigma]}(x/\xi) \,.
\end{equation}
In the Bjorken limit the Mandelstam variables for the hard-scattering
subprocess are
\begin{equation}
\hat{s} = x s + \half (1-x)\ms q^2 \,, \qquad
\hat{u} = x u + \half (1-x)\ms q^2 \,,
\end{equation}
so that one has $ x/\xi = (\hat{u} - \hat{s})/q^2$. To leading order
(LO) in $\alpha_s$, the kernel reads
\begin{equation}
  \label{born-kernel}
  C^{q [\sigma]}(\omega) \,\propto\,
  \frac{1}{1-\omega-i\epsilon} - \sigma \frac{1}{1+\omega-i\epsilon} \,,
\qquad
  \im C^{q [\sigma]}(\omega) \,\propto\, \pi\ms
  \bigl[\ms \delta(\omega-1) - \sigma \delta(\omega+1) \ms\bigr]
\hspace{1em}
\end{equation}
for both DVCS and meson production. At higher orders in $\alpha_s$ one
finds branch cuts in the $\hat{s}$ and $\hat{u}$ channels for
$\omega>1$ and $\omega<-1$, respectively. For the dispersion relations
we need to know the behavior of the kernels when $|\omega| \to
\infty$.  The NLO kernels for DVCS can be found in
\cite{Belitsky:2005qn}, and those for meson production in
\cite{Ivanov:2004zv}.  For negative signature, one finds $C^{q
  [-]}(\omega) \sim \omega^{-1}$ up to logarithms for both DVCS and
meson production.  For positive signature, the NLO corrections give
$C^{q [+]}(\omega) \sim \omega^{-1}$ for DVCS, and $C^{q[+]}(\omega)
\sim \omega^0$ for meson production, again up to logarithms.  The
power behavior as $\omega^0$ is due to two-gluon exchange in the
$t$-channel.  For DVCS such graphs only start at NNLO, so that at this
level one will also have $C^{q[+]}(\omega) \sim \omega^0$. For both
signatures one can thus write down an
unsubtracted dispersion relation for the kernel,
\begin{equation}
\re C^{q [\sigma]}\Bigl( \frac{x}{\xi} \Bigr) = \frac{1}{\pi}
\int_1^\infty d\omega\,
   \im C^{q [\sigma]}(\omega)
   \left[ \frac{1}{\omega - x/\xi} - \sigma \frac{1}{\omega + x/\xi}
   \right] \,.
\end{equation}
On the other hand, the invariant amplitude satisfies its own fixed-$t$
dispersion relation (\ref{disp-sub}). Therefore the real part of the
leading invariant amplitudes for DVCS or meson production can be
obtained from a dispersion relation for the hard-scattering kernel,
\begin{equation}
\re {\cal F}^{q [\sigma]}(\xi,t) =
  \frac{1}{\pi} \int_1^{\infty} d\omega\,
  \im C^{q [\sigma]}(\omega)\, \int_{-1}^1 dx\, F^q(x,\xi,t)
  \left[ \frac{1}{\omega\xi - x} -\sigma \frac{1}{\omega\xi + x}
  \right] \,,
  \label{hhh}
\end{equation}
or for the invariant amplitude itself,
\begin{eqnarray}
&& \re {\cal F}^{q [\sigma]}(\xi,t) =
  \frac{1}{\pi} \int_1^{\infty} d\omega\,
  \im C^{q [\sigma]}(\omega)\, \Biggl\{\,
  \int_{-1}^1 dx\, F^{q}\Bigl( x,\frac{x}{\omega},t \Bigr)
  \left[ \frac{1}{\omega\xi - x} -\sigma \frac{1}{\omega\xi + x}
  \right]
\hspace{2.8em}
\nonumber \\
&& \hspace{16em}
{} + {\cal I}^{q [\sigma]}(\omega,\xi_0,t) \Biggr\} \,,
  \label{disp-rel-fin}
\end{eqnarray}
where $\xi_0$ corresponds to the subtraction point $\nu_0$ in
(\ref{disp-sub}) and
\begin{equation}
  \label{I-def}
{\cal I}^{q [\sigma]}(\omega,\xi,t) = \int_{-1}^1 dx\,
  \biggl[ F^{q}(x,\xi,t)
        - F^{q}\Bigl(x,\frac{x}{\omega},t\Bigr) \biggr]
  \left[ \frac{1}{\omega\xi - x} - \sigma \frac{1}{\omega\xi + x}
  \right] \,.
\end{equation}
As shown in \cite{Diehl:2007jb}, the term ${\cal I}^{q [\sigma]}$ is
related with spin-zero exchange in the $t$-channel.

Consistency of the two representations provides nontrivial constraints
on the GPDs. Indeed, in (\ref{disp-rel-fin}) the GPD enters in the
DGLAP region only, whereas in (\ref{hhh}) both the DGLAP and ERBL
regions contribute. Let us see that the consistency is guaranteed by
the polynomiality property of Mellin moments, which follows directly
from the Lorentz covariance of the operator matrix elements that are
parameterized by GPDs.  With the conventional definitions (given e.g.\
in \cite{Diehl:2003ny}) we have for quarks
\begin{equation}
\label{pol}
\int_{-1}^1 dx\, x^{n-1}\, H^q(x,\xi,t) =
\sum_{k=0}^{n-1} (2\xi)^k\, A^q_{\smash{n,k}}(t)
 + (2\xi)^{n} C^q_{n}(t)\,,
\end{equation}
\begin{equation}
\label{pol1}
 \int_{-1}^1 dx\, x^{n-1}\, E^q(x,\xi,t) =
\sum_{k=0}^{n-1} (2\xi)^k\, B^q_{\smash{n,k}}(t)
 - (2\xi)^{n} C^q_{n}(t)\,,
\end{equation}
where $k$ is even because of time reversal invariance.  Clearly,
(\ref{hhh}) and (\ref{disp-rel-fin}) are consistent if ${\cal I}^{q
  [\sigma]}(\omega,\xi,t)$ is independent of $\xi$ for all $\omega \ge
1$.  To show that this is the case, we Taylor expand
$F^q(x,x/\omega,t)$ in its second argument,
\begin{eqnarray}
{\cal I}^{q [\sigma]}(\omega,\xi,t)
&=& \frac{1}{\omega}
\sum_{n=1}^\infty \frac{1}{n!}\,
   \Bigl( \frac{\partial}{\partial\eta} \Bigr)^n
   \int_{-1}^1 dx\,
     \Bigl(\frac{x}{\omega}-\xi\Bigr)^{n-1}
     F^q(x,\eta,t)\, \Bigr|_{\eta=\xi}
\nonumber \\
&+& \frac{\sigma}{\omega} \sum_{n=1}^\infty \frac{1}{n!}\,
   \Bigl( \frac{\partial}{\partial\eta} \Bigr)^n
   \int_{-1}^1 dx\,
     \Bigl(\frac{x}{\omega}+\xi\Bigr)^{n-1}
     F^q(x,\eta,t)\, \Bigr|_{\eta=-\xi} \,,
\end{eqnarray}
where we have interchanged the order of differentiation and
integration.  For definiteness let us consider the case $F^q = H^q$.
Using the polynomiality property (\ref{pol}) and the fact that $C^q_n$
is only nonzero for even $n$, we find
\begin{equation}
  \label{H-result-1}
{\cal I}^{q [+]}(\omega,\xi,t) = 2
\sum_{n=2}^\infty
    \left(\frac{2}{\omega}\right)^{n} C^q_{n}(t)\, ,
\qquad\qquad
{\cal I}^{q [-]}(\omega,\xi,t) = 0 \,,
\end{equation}
which is independent of $\xi$ as required.  In the case $F^q = E^q$
there is an additional minus sign on the r.h.s.\ of
(\ref{H-result-1}), in accordance with (\ref{pol1}).

The dispersion representations discussed here can provide a practical
check for GPD models in which Lorentz invariance is not exactly
satisfied.  In particular, we find that even for small $\xi$ the model
proposed in \cite{Freund:2002qf} leads to serious conflicts with
dispersion relations when it is used for calculating the real part of
scattering amplitudes \cite{Diehl:2007jb}.

The representation (\ref{disp-rel-fin}) has important consequences on
the information about GPDs that can be extracted from DVCS and meson
production.  To leading approximation in $\alpha_s$, the imaginary
part of the amplitude is only sensitive to the distributions at
$x=\xi$, and the only additional information contained in the real
part is a constant associated with pure spin-zero exchange, given by
(\ref{H-result-1}) at $\omega=1$.  In \cite{Teryaev:2005uj} this was
referred to as a holographic property. Beyond leading order, the
evaluation of both imaginary and real parts of the amplitude involves
the full DGLAP region $|x| \ge \xi$. In addition, the real part
depends on the appropriate spin-zero term at all $\omega \ge 1$.

Consider now the comparison of a given model or parameterization of
GPDs with data on DVCS or meson production.  In a leading-order
analysis (which should of course always be restricted to kinematics
where the LO approximation is adequate) it is sufficient to
characterize each GPD by its values at $x=\xi$, supplemented by a
constant for the spin-zero exchange contribution discussed above.  On
one hand this can be a welcome simplification, and on the other hand
it indicates the limitations of an LO analysis: when confronting data
with a given GPD one is sensitive to $x \neq\xi$ (and to the details
of the spin-zero exchange contribution) only at NLO or higher
accuracy.

\vspace*{0.8cm}

\noindent {\it This work is supported by the Helmholtz Association,
  contract number VH-NG-004. The work of D.I.\ is supported in part
  by the grants RFBR-06-02-16064 and NSh 5362.2006.2. }

\end{document}